\documentclass[9pt]{article}
\usepackage[latin1]{inputenc}
\usepackage{amssymb}
\usepackage{graphicx}
\begin{document}

\vspace{1.8cm}

\begin{center}
{\Large\bf Cosmologia neo-Newtoniana: um passo intermedi\'ario em dire\c{c}\~ao \`a Relatividade Geral \\[5PT]}
 {(\it Neo-Newtonian cosmology: An intermediate step towards General Relativity)}
\vspace{1.2cm}

J.C. Fabris$^{a,}$\footnote{e-mail: fabris@pq.cnpq.br} \medskip \vspace{0.5cm}
e\,\, H.E.S. Velten$^{b,}$\footnote{e-mail: velten@physik.uni-bielefeld.de} \medskip 
\par 
{ \it  a) Grupo de Gravita\c c\~ao e Cosmologia, 
Departamento de F\'{\i}sica, Universidade Federal do Esp\'{\i}rito Santo,
29060-900, Vit\'oria, Esp\'{\i}rito Santo, Brasil

b) Fakult\"at f\"ur Physik, Bielefeld Universit\"at, Postfach 100131, 33501 Bielefeld, Alemanha.
}

\abstract
Cosmology is a field of physics in which the use of General Relativity theory is indispensable. However, a cosmology based on Newtonian gravity theory for gravity is possible in certain circumstances. The applicability of Newtonian theory can be substantially extended if it is modified in such way that pressure has a more active role as source of the gravitational field. This was done in the neo-Newtonian cosmology. The limitation on the construction of a Newtonian cosmology, and the need for a relativistic theory in cosmology are reviewed. The neo-Newtonian proposal is presented, and its consequences for cosmology are discussed.
\vspace{0.3cm}


\abstract
A cosmologia \'e um dom\'{\i}nio da f\'{\i}sica no qual o emprego Teoria da Relatividade Geral \'e indispens\'avel. No entanto, uma cosmologia baseada na teoria Newtoniana da
Gravita\c{c}\~ao \'e poss\'{\i}vel em certas circunst\^ancias. O campo de aplicabilidade da teoria Newtoniana pode ser substancialmente estendido caso ela seja modificada de forma
a dar um papel mais ativo \`a press\~ao como fonte do campo gravitacional. Isto foi feito no \^ambito da teoria neo-Newtoniana. As limita\c{c}\~oes \`a constru\c{c}\~ao de uma cosmologia Newtoniana, e a
consequente necessidade de uma teoria relativista em cosmologia, s\~ao revistas. A proposta de uma teoria neo Newtoniana \'e apresentada, e sua consequ\^encias para a cosmologia s\~ao abordadas.
\newline
\vspace{0.3cm}
\newline
{\bf Palavras Chave:} Cosmologia Newtoniana, relatividade.
\vspace{0.5cm}
\end{center}

\section{Introdu\c{c}\~ao}

Por quase trezentos anos, as leis de Newton foram a melhor explica\c{c}\~ao para os
fen\^omenos mec\^anicos e gravitacionais, os fen\^omenos relativos ao movimento dos
corpos, incluindo os efeitos da
atra\c{c}\~ao gravitacional. Desde o "Principias"\, de Newton, a mec\^anica e a
gravita\c{c}\~ao Newtonianas reinaram de forma absoluta no que tange \`a
descri\c{c}\~ao daqueles fen\^omenos. Isso se deu at\'e o come\c{c}o do s\'eculo XX,
quando suas limita\c{c}\~oes ao descrever corretamente a mec\^anica de uma
part\'{\i}cula no limite de altas velocidades tornaram-se evidentes. \'E a partir
desse momento, que nos confrontamos com o advento das teorias relativistas, a
restrita formulada em 1905 \cite{einstein1905}, e a geral, formulada em 1915
\cite{einstein1916}. Estas duas teorias relativistas representaram uma
revolu\c{c}\~ao paradigm\'atica na F\'{\i}sica (no sentido empregado por Thomas Kuhn
\cite{kuhn}), a primeira no que se refere especificamente aos fen\^omenos
mec\^anicos, ao estabelecer a
exist\^encia de uma velocidade limite na natureza, e a segunda no que diz respeito
\`a gravita\c{c}\~ao, ao substituir a no\c{c}\~ao de for\c{c}a gravitacional pela de
curvatura do espa\c{c}o-tempo. Os novos paradigmas criados por estas teorias
relativistas determinariam uma nova dire\c{c}\~ao para toda a f\'{\i}sica, e isto
al\'em dos limites originais. O impacto das teorias relativistas, notadamente da
teoria geral da relatividade, foi particularmente determinante para a cosmologia. 
\par
Obviamente, a teoria da gravita\c{c}\~ao Newtoniana n\~ao era capaz de fazer as
mesmas predi\c{c}\~oes que a Relatividade Geral (RG), ao passo que a RG podia
reproduzir os resultados Newtonianos
no limite de campos fracos e baixas velocidades. Inevitavelmente, a cosmologia (onde
tal aproxima\c{c}\~ao de campo fraco e velocidades baixas dificilmente se aplica)
passa a ser associada com \`a complexidade matem\'atica existente na RG,
principalmente devido ao uso da geometria Riemanniana e suas no\c{c}\~oes correlatas
como a \'algebra tensorial, variedades diferenci\'aveis, etc. Estando a RG ligada a
uma geometria Riemanniana formulada em um espa\c{c}o-tempo quadri-dimensional,
parece ser a primeira vista imposs\'{\i}vel buscar o entendimento da din\^amica do
Universo sem todo esse aparato matem\'atico. O fato que a estrutura geom\'etrica na
qual a RG se baseia era quase que completamente estranha aos
outros dom\'{\i}nios da f\'{\i}sica, como, por exemplo, a ent\~ao emergente
mec\^anica qu\^antica, foi uma das raz\~oes do relativo ostracismo que essa nova
teoria da gravita\c{c}\~ao viveu entre os anos 30 e os anos 60, per\'{\i}odo que foi
denominado de {\it travessia do deserto} por Einsenstaedt \cite{jean}.
\par
\'E natural esperar que teorias f\'isicas mais complexas sejam precedidas por
formula\c{c}\~oes mais elementares e n\~ao t\~ao pretensiosas. Ocorre que, na
d\'ecada de 1930, os trabalhos de E. A. Milne e W. H. McCrea mostraram que a
cosmologia poderia ser tratada n\~ao somente atrav\'es da complexidade matem\'atica
e conceitual inserida na RG, mas tamb\'em por uma formula\c{c}\~ao mais simples e
elementar como a que encontramos na f\'{\i}sica Newtoniana \cite{milne,milnemccrea}.
Este \'e, talvez, um dos raros epis\'odios na hist\'oria da ci\^encia onde ocorre
uma invers\~ao cronol\'ogica no desenvolvimento de alguma expertise. Desta forma,
surge a possibilidade de utilizar um tratamento Newtoniano, de grande simplicidade
matem\'atica, para problemas at\'e ent\~ao delegados somente \`a cosmologia
relativista. De fato, como veremos na pr\'oxima se\c{c}\~ao, a cosmologia Newtoniana
\'e capaz de descrever a evolu\c{c}\~ao do Universo durante toda a fase
 dominada pela mat\'eria (definida pela condi\c{c}\~ao de press\~ao nula, $p=0$)
onde as estruturas c\'osmicas como gal\'axias e aglomerados se formaram. 
\par
Apesar da formula\c{c}\~ao de Milne e McCrea ser suficiente para analisar muitos
problemas dentro da cosmologia, havia uma grandeza fundamental que at\'e ent\~ao
n\~ao era considerada: a press\~ao $\left( p \right)$. A partir da RG sabemos que a
press\~ao possui papel fundamental na din\^amica do Universo pois \'e capaz de
induzir efeitos gravitationais assim como a pr\'opria mat\'eria. A press\~ao "pesa"
e, portanto, gravita. Para entender como a press\~ao influi na din\^amica do
Universo basta olharmos as equa\c{c}\~oes apresentadas abaixo, obtidas pelo
matem\'atico russo Alexander Friedmann em 1922 a partir das equa\c{c}\~oes de
Einstein, e que s\~ao as equa\c{c}\~oes base da cosmologia relativista
\cite{friedmann}.
\par
A teoria da Relatividade Geral \'e definida pelas equa\c{c}\~oes de Einstein:
\begin{eqnarray}
R_{\mu\nu} - \frac{1}{2}g_{\mu\nu}R &=& 8\pi GT_{\mu\nu},\\
{T^{\mu\nu}}_{;\mu} &=& 0. \label{cons}
\end{eqnarray}
onde $R_{\mu\nu}$ \'e o tensor de Ricci dado por $R_{\mu\nu} = \partial_\rho
\Gamma^\rho_{\mu\nu} - \partial_\nu \Gamma^\rho_{\mu\rho} +
\Gamma^\rho_{\mu\nu}\Gamma^\sigma_{\rho\sigma}
- \Gamma^\rho_{\mu\nu}\Gamma^\sigma_{\rho\sigma}$, sendo $\Gamma^\rho_{\mu\nu} =
\frac{1}{2}g^{\rho\sigma}(\partial_\mu g_{\sigma\nu} + \partial_\nu g_{\sigma\mu} -
\partial_\sigma g_{\mu\nu})$,
e $R = g^{\rho\sigma}R_{\rho\sigma}$ \'e o escalar de Ricci. Os \'{\i}ndices $\mu,
\nu, \rho,...$ designam  as coordenadas espa\c{c}o-temporais. O subscrito ";" indica
derivada covariante em uma
geometria curva. S\~ao usadas nestas express\~oes
a conven\c{c}\~ao de que \'{\i}ndices repetidos implicam uma soma. A m\'etrica
$g_{\mu\nu}$ define a geometria,
atrav\'es da express\~ao da dist\^ancia infinitesimal entre dois pontos no
espa\c{c}o-tempo:
\begin{equation}
ds^2 = g_{\mu\nu}dx^\mu dx^\nu.
\end{equation}
O termo $T_{\mu\nu}$ define o conte\'udo de mat\'eria e energia. Para um fluido
perfeito, esse {\it tensor de momento-energia} assume a forma,
\begin{equation}
T_{\mu\nu} = (\rho + p/c^2)u^\mu u^\nu - p g_{\mu\nu}.
\end{equation}
Logo, as equa\c{c}\~oes de Einstein conectam a geometria do espa\c{c}o-tempo com a
distribui\c{c}\~ao de mat\'eria e energia.
\par
A geometria que descreve um Universo homog\^eneo, isotr\'opico e que se expande \'e
definida pela m\'etrica de Friedmann-Lema\^{\i}tre-Robertson-Walker (FLRW):
\begin{equation}
ds^2 = dt ^2 - a^2(t)\biggr\{\frac{dr^2}{1 - kr^2} + r^2(d\theta^2 +
\mbox{sen}^2\theta d\phi^2)\biggl\},
\end{equation}
onde o par\^ametro $k$ indica se a se\c{c}\~ao espacial a tempo constante \'e o
espa\c{c}o euclideano ($k = 0$), uma tri-esfera ($k = 1$), ou uma tri-pseudo esfera
($k = - 1$).
Com esta m\'etrica, a componente $0-0$ da equa\c{c}\~ao de Einstein fornece:
\begin{equation}
\label{friedmann1}\frac{\dot{a}^{2}}{a^{2}}+\frac{k\,c^2}{a^{2}}=\frac{8 \pi G
\rho}{3},
\end{equation}
onde $a$ \'e fator de escala do Universo (que define suas dimens\~oes, por exemplo,
ao raio da tri-esfera para o caso $k = 1$), $k$ est\'a associado, como foi dito, \`a
curvatura do Universo, $G$ \'e a constante universal da gravita\c{c}\~ao, $c$ \'e a
velocidade da luz e $\rho$ \'e densidade de mat\'eria-energia do Universo. 
\par
Ao mesmo tempo, a componente $i-i$ desta equa\c{c}\~ao fornece uma express\~ao para
a segunda derivada do fator de escala:
\begin{equation}
\label{friedmann2}\frac{2\ddot{a}}{a}+\frac{\dot{a}^{2}}{a^{2}}+\frac{k\,c^{2}}{a^{2}}=-\frac{8
\pi G p}{c^{2}}.
\end{equation}
As equa\c c\~oes (\ref{friedmann1}) e (\ref{friedmann2}) podem ser combinadas,
resultando em:
\begin{equation}
\label{friedmann3}\frac{\ddot{a}}{a}=- \frac{4 \pi G}{3}\left( \rho + \frac{3
p}{c^{2}} \right).
\end{equation}
A equa\c{c}\~ao (\ref{friedmann3}) fornece uma express\~ao para a acelera\c{c}\~ao
do Universo em termos da densidade $\rho$ e da press\~ao $p$, que s\~ao as
componentes do tensor momento-energia, respons\'avel por descrever as
caracter\'{\i}sticas f\'{\i}sicas das distintas esp\'ecies de componentes de
mat\'eria e energia que preenchem o Cosmo (neutrinos, f\'otons, b\'arions, etc.).
Uma outra informa\c{c}\~ao \'e necess\'aria para fechar este conjunto de equa\c
c\~oes e a obtemos, em geral, assumindo que a mat\'eria constituinte do Universo \'e
caracterizada por uma equa\c{c}\~ao de estado $p=p\left( \rho \right)$. Inserindo
essa equa\c{c}\~ao de estado na equa\c{c}\~ao de conserva\c{c}\~ao 
\begin{equation}
\dot{\rho}+3\frac{\dot{a}}{a}(\rho+p)=0,
\end{equation}
que \'e a componente $0-0$ da equa\c{c}\~ao (\ref{cons}),
obtemos o comportamento da densidade e, por conseguinte, todas as demais solu\c
c\~oes para a din\^amica do universo.
\par
Com estes resultados, Friedmann observou a possibilidade de se obter um Universo
din\~amico, onde seu comportamento dependeria unicamente de sua distribui\c{c}\~ao
de mat\'eria e energia. At\'e ent\~ao, o conceito de um Universo est\'atico ainda
era fortemente defendido, inclusive por Einstein, que, com essa cren\c{c}a, havia
introduzido em suas equa\c c\~oes um termo chamado constante cosmol\'ogica
$\Lambda$. No entanto, em 1929, Edwin Hubble, trabalhando no observat\'orio do Monte
Wilson, observou que as gal\'axias se afastavam umas das outras com uma velocidade
proporcional \`a sua dist\^ancia. Hubble teria acabado de demonstrar que o Universo,
na verdade, est\'a experimentando uma fase de expans\~ao, ou seja, \'e din\^amico
\cite{hubble}. As descobertas de Hubble j\'a tinham sido, de certa forma,
antecipadas por Lema\^{\i}tre \cite{lemaitre}.
O cr\'edito \`a descoberta da expans\~ao do Universo \'e, hoje, objeto de
discuss\~ao \cite{bergh}.

\par
A constante cosmológica possui sua origem ligada à concepção de Einstein de que o Universo deveria ser estático. No entanto, grande parte das observa\c c\~oes astronômicas, incluindo, principalmente, supernovas tipo Ia, indicam que o universo, já há aproximadamente 6 bilh\~oes de anos, experimenta uma fase de expansão acelerada possivelmente causada por $\Lambda$. Na verdade, os indícios observacionais nos levam a acreditar que essa misteriosa constante $\Lambda$ contribui com $70\%$ da energia do Universo representando o que chamamos de energia escura. Os $30\%$ restantes estariam divididos sob a forma de matéria escura $(25\%)$ e matéria bariônica $(5\%)$. Este cenário é conhecido como modelo cosmológico $\Lambda$CDM e figura como dinâmica padrão do Universo.
\par 
Em princ\'{\i}pio, o modelo $\Lambda$CDM requer, para sua correta descri\c{c}\~ao, a RG. De uma forma mais geral, em qualquer modelo onde a press\~ao do constituinte deve desempenhar um
papel din\^amico, requereria a RG. No entanto, \'e poss\'{\i}vel modificar a teoria Newtoniana usual de maneira que a press\~ao assuma um papel relevante, mesmo em situa\c{c}\~oes onde se sup\~oe
que o conte\'udo de mat\'eria e energia \'e homog\^eneo. Desta forma, modelos cosmol\'ogicos que guardam as caracter\'{\i}sticas essenciais ditados pela teoria relativista podem ser constru\'{\i}dos
no \^ambito de uma teoria Newtoniana modificada. Isto implica um ganho excepcional em simplicidade matem\'atica e conceitual.
\par
Nosso objetivo aqui será expor o caminho percorrido pela cosmologia que levou ao que chamamos hoje de "Cosmologia (neo-)Newtoniana"'. Na próxima seção, apresentamos as primeiras equa\c c\~oes Newtonianas, desenvolvidas por Milne e McCrea, em 1934, para um Universo preenchido por matéria não relativística $(p=0)$. Esse fluido dominou o Universo desde que sua idade era de aproximadamente $300.000$ anos até ``muito recentemente'', quando a energia escura passa a dominar a dinâmica cósmica. Na terceira seção mostramos como E. R. Harrison, na década de 1960, conseguiu incorporar a pressão nas equa\c c\~oes da cosmologia Newtoniana obtidas anteriormente. Essa modifica\c{c}\~ao foi revisada e aprimorada posteriormente na refer\^encia \cite{ademir}. Temos assim a chamada cosmologia neo-Newtoniana e com isso, passamos a ter domínio da cosmologia durante os primeiros 300.000 anos da história do Universo, a época dominada pela radiação ($p=\rho/3$), que é imediatamente anterior a fase da matéria ($p=0$). A última e atual fase, dominada pela energia escura, também se beneficiará da cosmologia neo-Newtoniana. Como em uma fase de expansão acelerada temos $\ddot{a}>0$, as equa\c c\~oes de Friedmann nos dizem que a energia escura deve apresentar uma pressão negativa (veja equação \ref{friedmann3}).
\par
Para uma série de trabalhos sobre a teoria relativista recomendamos ao leitor a edição I/2005 da RBEF dedicada aos cem anos da relatividade. Em especial, a evolução da cosmologia nestes cem anos estão expostas em \cite{waga}. Uma importante discuss\~ao sobre a cosmologia Newtoniana também aparece em \cite{marcelo}.

\section{Cosmologia Newtoniana}

O primeiro trabalho realizado com o propósito de buscar uma alternativa para a cosmologia relativista foi feito por E. A. Milne em 1934 \cite{milne}. Nele foi pela primeira vez observada a possibilidade de uma abordagem distinta da relativista para a cosmologia. Em 1929, as observações de Hubble forneceram fortes evidências em favor da expansão do Universo. O mecanismo relativista para explicar as observações de Hubble considerava que o Universo, como um todo, possuia uma dinâmica associada à expansão. Quer dizer, todo o Universo relativístico expande. Por outro lado, a formulação proposta por Milne remetia a um Universo estático, Newtoniano, onde a expansão observada é devida aos movimentos de partículas (galáxias, estrelas, etc.) neste espaço estático - em uma linguagem mais t\'ecnica, as obseva\c{c}\~oes da
expans\~ao seriam devido ao campo peculiar de velocidades. Com isso, partículas se afastando umas das outras em um Universo estático (Newtoniano) forneceriam a mesma observação que partículas imóveis em um Universo que se expande (relativístico). Ao contrário da RG, a proposta de Milne preteria uma modificação na geometria, afim de manter uma geometria Euclideana. A principal vantagem apontada por Milne para essa abordagem, era a utilização do espaço-tempo de Minkowski, que \'e o espa\c{c}o-tempo da Relatividade Restrita e, por extens\~ao, da Teoria de Campos, cl\'assica e qu\^antica. Esta afirmação de Milne é fundamentada pelas palavras do próprio Albert Einstein\cite{Relatividade}: {\it ''Pode-se considerar o mundo de Minkowski, do ponto de vista formal, como um espaço euclidiano quadridimensional (com coordenada temporal imaginária)''}. Por outro lado, a grande diferença entre a cosmologia Newtoniana e relativista é a interpretação das quantidades físicas que aparecem em suas equações. Na cosmologia Newtoniana o tempo $t$ é um tempo absoluto, enquanto que, o tempo $\tau$, relativístico, é o tempo cósmico (tempo medido por um observador co-móvel). No entanto, exceto essa questão, vale-se ressaltar que cosmologias relativista e Newtoniana predizem, localmente, os mesmos resultados. Esse fato nos leva a uma questão: por que a teoria Newtoniana e a relatividade geral, quando aplicadas a um Universo uniforme, levam aos mesmos resultados?
\par
Em um trabalho seguinte \cite{milnemccrea} Milne e McCrea fundamentaram o que, posteriormente, viria a ser chamado de Cosmologia Newtoniana \cite{layzer}. Este trabalho é, na verdade, uma generalização do anterior. No entanto, seu grande mérito foi a obtenção da equação de Friedmann (\ref{friedmann1}-\ref{friedmann3}), para o fator de escala do Universo, a partir de um tratamento puramente Newtoniano. Assim, utilizando apenas as leis da dinâmica e gravitação Newtoniana eles mostraram que as equações relativistas poderiam ser obtidas através de uma abordagem muito mais simples. Para isso, as únicas considerações utilizadas foram o Princípio Cosmológico, caracterizando a homogeneidade e isotropia do Universo e a exigência de que a pressão é pequena o suficiente, quando comparada à densidade, para que pudesse ser desprezada. Dessa forma, Milne e McCrea encontraram que a equação Newtoniana, responsável por reger a evolução de uma partícula de massa $m$ e energia total $E$ localizada a uma distância $R$ do centro de uma esfera homogênea e isotrópica, constituída de matéria com densidade $\rho \left( t \right)$ é:
\begin{equation}
\label{milne1}\frac{\dot{R}^{2}}{R^{2}}+\frac{\left( \frac{-2E}{m}\right)}{R^{2}}=\frac{8\pi G \rho}{3}\,.
\end{equation}
\par
Este resultado possui uma íntima relação com a segunda lei de Newton. Para isso, basta considerar uma partícula de massa $m$ situada a uma distância $R$ do centro de uma distribuição esférica de matéria, com massa $M$ interior a $R$, caracterizada por uma densidade $\rho=3M/4\pi R^{3}$. Naturalmente, as leis de Newton tratam este problema como se toda massa $M$ estivesse concentrada no centro dessa distribuição. Assim, a força Newtoniana exercida sobre a massa $m$ é:
\begin{equation}
m \frac{d^{2} R}{dt^{2}}=-\frac{GMm}{R^{2}}\,,
\end{equation}
que coincide com a equação relativística para a evolução da matéria ($p$=0) \footnote{Mais tarde se ver\'a que n\~ao \'e necess\'ario considerar $p = 0$ e sim que a press\~ao \'e homog\^enea}. Se usarmos $M=\frac{4\pi R^{3}\rho}{3}$ teremos,
\begin{equation}
\label{milne3}\ddot{R}=-\frac{4\pi G \rho R}{3}\,.
\end{equation}
Se assumimos que a densidade varia de acordo com $\rho=\rho_{0}\frac{R^{3}_{0}}{R^{3}}$, onde $R_{0}$ é uma constante, e multiplicarmos a equação (\ref{milne3}) por $\dot{R}$, o resultado pode ser integrado obtendo-se uma equação formalmente idêntica à equação (\ref{friedmann1}). Porém, a constante de integração que surge neste procedimento possui uma interpretação distinta ao seu análogo $k$ na equação (\ref{friedmann1}). Na teoria Newtoniana, esta quantidade \'e a energia por unidade de massa do sistema.
\par
O grande resultado desses trabalhos foi mostrar que as predições locais das cosmologias Newtonianas e relativísticas são exatamente as mesmas, já que muitos de seus resultados básicos são algebricamente equivalentes, quando $p=0$. Com isso a cosmologia Newtoniana torna-se útil como uma primeira aproximação para a cosmologia, para só então, em uma etapa posterior, se fazer necessário o uso da RG.
Para se ter uma ideia de sua utilidade, as  simula\c c\~oes numéricas que tentam reproduzir, com auxílio de super-computadores, a distribui\c c\~ao de massa do Universo utilizam o modelo newtoniano (veja http://www.deus-consortium.org/ e http://www.mpa-garching.mpg.de/galform/millennium/).
\par
A similaridade entre a equação (\ref{milne1}) e a equação (\ref{friedmann1}) é notável. Se considerarmos $\frac{-2E}{m}=k$, percebe-se que essas equações são idênticas. Contudo, essa correspondência implica em uma série de consequências, de natureza teórica e interpretativa, que criam divergências sobre esta dedução das equações cosmológicas.
\par
Um dos primeiros críticos da idéia de uma abordagem Newtoniana para o problema cosmológico foi David Layzer. Em 1954, Layzer, após a demonstração de dois teoremas, conclui que a teoria de Milne e McCrea possui limitações pricipalmente ao tratar de sistemas não ligados. Além disso, a cosmologia Newtoniana, assim como proposta por Milne e McCrea, trata apenas de distribuições finitas de matéria em um espaço euclideano infinito com matéria distribu\'{\i}da de maneira homogênea e isotrópica \cite{layzer}. Por outro lado, os defensores da cosmologia Newtoniana, incluindo o próprio McCrea\footnote{No entanto, não mais Milne que havia falecido em 1950.}, baseavam-se em argumentos atribu\'{\i}dos a H. Bondi \cite{bondi}, que afirmavam que a cosmologia Newtoniana poderia ser utilizada em regiões finitas do Universo formando um sistema isolado, o que tornaria viável o uso da cosmologia Newtoniana, ao menos para as aplicações vislumbradas na época. Em sistemas isolados, alguns dos problemas principais associados a uma distribui\c{c}\~ao de mat\'eria que se estende ao infinito podem ser contornados.
\par
De fato, podemos citar duas dificuldades conceituais principais relativas \`a constru\c{c}\~ao de uma cosmologia newtoniana. Tal cosmologia implica, se se usa o princ\'{\i}pio cosmol\'ogico, que 
a mat\'eria deve se distribuir de forma homog\^enea e infinita. Em tal circunst\^ancia, como o campo gravitacional se estende tamb\'em infinitamente, n\~ao h\'a como definir um referencial inercial
onde a segunda lei seria aplicada. Em segundo lugar, em uma distribui\c{c}\~ao infinita, se um ponto de refer\^encia qualquer \'e escolhido e tenta-se calcular, usando a lei de Gauss, a for\c{c}a gravitacional sobre uma part\'{\i}cula teste situada a uma dist\^ancia $\vec r$ daquele ponto arbitr\'ario, a intensidade da for\c{c}a sobre a part\'{\i}cula teste, assim como a sua dire\c{c}\~ao, dependem
do ponto central escolhido. Tal dificuldade parece estar ligada ao problema do uso da lei de Gauss em um sistema onde o campo n\~ao se anula no infinito, e onde a defini\c{c}\~ao de sistema inercial
n\~ao \'e poss\'{\i}vel. Na refer\^encia \cite{mavrides} uma solu\c{c}\~ao apontada foi a admitir que a defini\c{c}\~ao de sistema inercial \'e poss\'{\i}vel apenas localmente, utilizando a no\c{c}\~ao de
queda livre \footnote{{\it Le syst\`eme de chaque observateur est inertiel, localement, mais les diff\'erents observateurs peuvent avoir un mouvement acc\'eler\'e les uns par rapport aux autres. Ceci n'est pas admissible au sens strict de la M\'ecanique Classique
mais, aussi longtemps que chaque observateur n'utilise que son propre syst\`eme, aucune difficult\'e ne surgit} \cite{mavrides}.}.  
\par
Destaca-se ainda, no entanto, uma segunda volta a este debate situada no fim do século $XX$. Em uma série de artigos voltados para a filosofia da ciência, John Norton revive essa discussão alegando que a teoria Newtoniana determina, de maneira inconsistente, a força gravitacional quando aplicada a uma distribuição homogênea de massa \cite{norton1993}. Seu argumento tem como base a representação integral para a força gravitacional $\int G\rho\left(r^{\prime}\right) \left({\bf r}-{\bf r}^{\prime}\right)\left|{\bf r}-{\bf r}^{\prime}\right|^{-3} d{\bf V^{\prime}}$, que é geralmente encontrada nos livros texto - se a densidade se mantém finita em uma distribui\c{c}\~ao homog\^enea que se estende at\'e o infinito, esta integral se torna divergente e o campo gravitacional $\vec{g}$ dependeria das condi\c c\~oes de contorno no infinito. Como ilustra\c c\~ao, se imaginarmos uma esfera de raio $R$, o teorema de Gauss fornece $\vec{g}=-(4\pi/3)G \rho \vec{r}$ para $r<R$. Este resultado n\~ao muda se $R\rightarrow \infty $, logo, concluímos que  $\vec{g}$ é bem definido em qualquer $r$ finito. Suponha, no entanto, uma distribui\c c\~ao esferóide (uma esfera alongada, de certa ecentricidade $\epsilon$). A única diferen\c ca na distribui\c c\~ao de massa seria na camada entre o esferóide e sua esfera circuscrita e, logo, $\vec{g}$ mudaria, exceto no centro desse sistema. Com isso, o campo gravitacional depende das condi\c c\~oes de contorno impostas no infinito. Como solução, Norton sugere modificar a teoria da gravitação Newtoniana, impondo o que ele chama de "`relatividade da aceleração"'\cite{norton1995}. Por outro lado, David Malament \cite{malament} observa que as dificuldades apontadas por Norton são artefatos da formulação integral e desaparecem se passamos a considerar uma formulação "`geometrizada"' da gravitação. Por formulação geometrizada, Malament refere-se ao uso de técnicas introduzidas por Cartan \cite {cartan} e Friedrichs \cite{friedrichs} nos anos de 1920, para reconstruir uma teoria da gravitação que possui resultados observacionais idênticos aos da teoria original Newtoniana. Uma moderna revisão desse formalismo pode ser encontrado em \cite{tipler}. Contudo, mesmo em meio a críticas e sucessos, a teoria de Milne e McCrea figurou durante o século 20 como importante aproximação para o problema cosmológico.
\par
O espaço da cosmologia Newtoniana é estático, logo, é o movimento de partículas neste espaço que promove a expansão do Universo observada. Dessa forma, é necessário descrever o movimento dessas partículas através de algum conjunto de equações. A homogeneidade e isotropia do Universo motivam o chamado Princípio Cosmológico. Uma das principais observações que indicam esta característica é a Radiação Cósmica de Fundo. Além disso, em grandes escalas, a distribuição de matéria praticamente não apresenta flutuações (não homogeneidades), o que indica que podemos tratar o Universo como se fosse preenchido por um fluido, assumindo assim a hipótese do contínuo. Segundo esta hipótese, para atribuir o caráter de fluido à uma substância, é necessário que o menor elemento de volume considerado contenha um número suficiente de "partículas" \,para que as propriedades médias da substância variem de maneira contínua. No caso de assumirmos que o Universo possui este comportamento, as equações que são utilizadas para descrever o movimento desse fluido são as equações da hidrodinâmica usual \cite{landau}. Com isso, a dinâmica das partículas constituintes do Universo, caracterizada por uma interação gravitacional, pode ser descrita pelas seguintes equações:
\par
1) da continuidade (conserva\c{c}\~ao da mat\'eria),
\begin{equation}
\label{e1}
\label{continuidade}\frac{\partial \rho}{\partial t}+  \vec \nabla.(\rho  \vec u)=0\,,
\end{equation}
\par
2) de Euler (segunda lei de Newton, incluindo o efeito da press\~ao e do campo gravitacional, escrita em coordenadas Eulerianas),
\begin{equation}
\label{e2}
\label{euler}\frac{\partial \vec u}{\partial t}+ \left(\vec u .\vec \nabla \right)\vec u= -\vec \nabla \Psi - \frac{\vec \nabla p}{\rho}\,,
\end{equation} 
\par
3) de Poisson (lei da gravita\c{c}\~ao Newtoniana),
\begin{equation}
\label{e3}
\label{poisson}\nabla^{2}\Psi=4 \pi G \rho\,,
\end{equation}
onde $\rho$ é a densidade do fluido, $\vec u$ é o campo de velocidades, $\Psi$ é o potencial gravitacional e $p$ a pressão do fluido. A pressão é descrita por uma equação do tipo $p=p\left( \rho \right)$, denominada equação de estado do fluido. Estas s\~ao as equa\c{c}\~oes de um fluido ordin\'ario na presen\c{c}a do campo de gravita\c{c}\~ao \cite{landau}. Um modelo cosmol\'ogico pode ser constru\'{\i}do, incorporando a expans\~ao do universo, a homogeneidade e a isotropia, supondo
\begin{equation}
\rho = \rho(t), \quad p = p(t) \,\,\mbox{e} \quad \vec v = \frac{\dot a}{a}\vec r,
\end{equation}
onde $a = a(t)$ \'e uma fun\c{c}\~ao do tempo, que define a escala do Universo. A inser\c{c}\~ao
destas express\~oes nas equa\c{c}\~oes (\ref{e1},\ref{e2},\ref{e3}) conduz \`a equa\c{c}\~ao de Friedmann. Observe-se que, como nas equa\c{c}\~oes acima aparece apenas o gradiente espacial da press\~ao,
uma press\~ao suposta homog\^enea n\~ao contribui para a din\^amica do sistema.

\section{Cosmologia neo-Newtoniana}

Como já exposto, o trabalho realizado por Milne e McCrea em 1934 mostrou que as equações relativísticas para a dinâmica do Universo poderiam ser obtidas através de um tratamento Newtoniano considerando um caso onde a pressão é nula. Naturalmente, o próximo passo no aprimoramento de uma cosmologia Newtoniana seria considerar a pressão desse suposto fluido cósmico. Esta generalização foi feita pelo prórpio McCrea em 1951 \cite{mccrea1951} e posteriormente aperfeiçoada por E. R. Harrison em 1965 \cite{harrison}, resultando no que chamamos de cosmologia neo-Newtoniana.
\par
McCrea mostrou que, afim de se fazer prevalecer a analogia entre os casos Newtoniano e relativista, é necessário a adoção de dois conceitos físicos da relatividade. Primeiro, manter a equivalência entre massa e energia, através do fator $c^{2}$. Além disso, assumir a possibilidade de distinção entre massa inercial e gravitacional. Isso porquê, seria fundamental considerar que a densidade de massa gravitacional $\sigma$ de uma distribuição de matéria e energia era fornecida por \cite{whittaker}:
\begin{equation}
\sigma=T^{0}_{\,\,0}-T^{1}_{\,\,1}-T^{2}_{\,\,2}-T^{3}_{\,\,3}.
\end{equation}
Onde $T^{i}_{\,\,i}$ são as componentes diagonais do tensor momento-energia presente na equação de Einstein e definido anteriormente. Essas componentes são escritas como:
\begin{equation}
T^{0}_{\,\,0}=\rho\hspace{1cm} T^{1}_{\,\,1}=T^{2}_{\,\,2}=T^{3}_{\,\,3}=-\frac{p}{c^{2}}\hspace{1cm} T^{\mu}_{\,\,\nu}=0\rightarrow(\mu\neq\nu).
\end{equation}
Nas express\~oes acima, utilizou-se que $u^0 = u_0 = 1$, $u^i = u_i = 0$ e que $g^\mu_\nu = \delta_\nu^\mu$.
Com isso, a densidade de massa gravitacional é escrita por:
\begin{equation}
\sigma=\rho+\frac{3p}{c^{2}}.
\label{densidade}
\end{equation}
Esta combinação é a que determina, na Relatividade Geral, a convergência das
geodésicas. Sua positividade corresponde à condição de energia forte \cite{Ellis}. 
Para um Universo preenchido por um fluido de densidade $\rho$ e pressão $p$,  em expansão, e considerando que um observador em um ponto $\emph{O}$ descreve a velocidade de um objeto na posição $\vec q$ em relação a $\emph{O}$ como $d \vec q /dt= H(t) \,\vec q$, sua aceleração é fornecida por:
\begin{equation}
\label{acel}\frac{d^{2}\vec q}{dt^{2}}= \vec q \hspace{.1cm}\frac{d H}{dt}+\frac{d\vec q}{dt}\hspace{.1cm}H=\vec q\hspace{.1cm}\dot{H}+\vec q \hspace{.1cm}H^{2}.
\end{equation}  
Se o observador em $\emph{O}$ vê todo o espaço ao seu redor esfericamente simétrico então o módulo da força gravitacional exercida sobre o objeto em $\vec q$ pela massa contida no interior de uma esfera de raio $q$ será $GM/q^{2}$. É neste ponto que deve-se considerar que a densidade de massa é fornecidada por $\sigma$ \cite{whittaker}. Assim, quando considerarmos a aceleração da partícula, equação (\ref{acel}), devida à força gravitacional, teremos:
\begin{equation}
\dot{H}+H^{2}=-\frac{4}{3}\pi G \sigma=-\frac{4}{3}\pi G \left( \rho+\frac{3p}{c^{2}}\right).
\label{neoNewtace}
\end{equation}
Quando identificamos $H \left( t \right)$ com o fator de escala do universo através da lei de Hubble, $H \left( t \right)=\frac{\dot{a}}{a}$, encontramos a partir da equação acima as equações de Friedmann descritas na primeira seção. 
Dessa maneira, McCrea conseguiu coincidir suas novas equações diferenciais, obtidas para o fator de escala do Universo, com as equações de Friedmann para o caso com pressão não nula, apenas considerando uma simetria esférica e um Universo homogêneo e isotrópico.
\par
Posteriormente, em 1965, Harrison, sem utilizar qualquer conceito proveniente da relatividade, obtém os mesmos resultados encontrados por McCrea em 1951. Harrison utilizou o conceito de microcosmo, onde o Universo seria particionado em células de volume infinitamente pequeno imersas em um espaço-tempo plano. Estas células, não importando o quão pequenas sejam, estão em cada instante com o mesmo conteúdo, ou seja, são indistinguíveis. O estado de cada célula é independente do seu volume. Com isso, quando fazemos o volume da célula tender a zero, a métrica no interior da célula permanece plana como na relatividade especial. Assim, como o Princípio Cosmológico aplicado a um fluido perfeito garante a homogeneidade e isotropia do Universo, pode se dizer o mesmo sobre o conceito de microcosmo de Harrison. O que Harrison concluiu foi que o modelo de microcosmo não necessita de conceitos advindos da Relatividade Geral. Dessa forma, utilizando apenas o formalismo da relatividade especial associado à primeira lei da termodinâmica e as equações hidrodinâmicas, Harrison obteve resultados idênticos aos de McCrea.
\par
Como resultado dos trabalhos de Harrison e McCrea, obtemos a cosmologia neo-Newtoniana, cujas equações hidrodinâmicas são as seguintes \cite{harrison}:
\par
1) equação da continuidade\,,
\begin{equation}
\label{continuidade2}\frac{\partial \rho}{\partial t}+ \vec\nabla .\left[\left( \rho + \frac{p}{c^{2}}\right)\vec u\right]=0\,,
\end{equation}
\par
2) equação de Euler\,,
\begin{equation}
\label{euler2}\frac{\partial \vec u}{\partial t}+ \vec u . \vec\nabla \vec u = - \vec\nabla \Psi - \left(\rho +\frac{p}{c^{2}}\right)^{-1}\vec \nabla p\,,
\end{equation}
\par
3) equação de Poisson\,,
\begin{equation}
\label{poisson2}\nabla^{2} \Psi = 4 \pi G \left( \rho + \frac{3p}{c^{2}}\right).
\end{equation}
Na equação da continuidade e de Euler, considera-se que a corrente de matéria é
dada por $(\rho + p/c^2) \vec v$. Esta identificação esta associada ao fluxo de
energia. Por exemplo, em uma teoria relativista a positividade
de $\rho + p/c^2$ assegura que esse fluxo de energia esta conectado a um vetor do
tipo tempo, sem violação da causalidade \cite{Ellis}. Nestas equações, todas as noções da física Newtoniana são mantidas, como por exemplo, tempo absoluto, espaço Euclideano e força gravitacional. Neste conjunto de equações a pressão possui um papel muito mais notável. A densidade de massa gravitacional ($\sigma$) que compõe a equação de Poisson é fornecida pela equação (\ref{densidade}). As soluções das equações (\ref{continuidade2}-\ref{poisson2}) fornecem a descrição da evolução da dinâmica do Universo na presença de um fluido perfeito com pressão $p$. Entretanto, em um trabalho mais recente, Lima et. al. \cite{ademir} mostraram que essas equações não são satisfatórias a nível perturbativo. Quando introduzimos pequenas não homogeneidades nessas equações, com o objetivo de estudar o processo de formação das estruturas cósmicas como galáxias e aglomerados, não é possível obter a equação análoga à obtida na teoria relativista. Uma vez que a teoria relativista prediz o crescimento observado das estruturas cósmicas, obviamente, isso seria um forte argumento contra o uso da cosmologia neo-Newtoniana. Estes autores resolveram este problema redefinindo a equação da continuidade (\ref{continuidade2}) através de argumentos termodinâmicos. Assim, a forma correta para a equação da continuidade na cosmologia neo-Newtoniana é, definitivamente
\begin{equation}
\frac{\partial \rho}{\partial t}+ \vec\nabla .\left( \rho\vec u\right) + \frac{p}{c^{2}}\vec\nabla\vec u=0.
\end{equation}
No entanto, note que esta foi a forma utilizada originalmente por McCrea. O último termo dessa equação está relacionado ao trabalho ($d\tau =p dV$) necessário para expandir uma esfera de um volume $V$ até $V+dV$:
\begin{equation}
\frac{1}{V}\frac{d\tau}{dt}=p\frac{4 \pi a^{2} da}{\frac{4}{3}\pi a^{3}dt}=3\frac{\dot{a}}{a}p=p \vec\nabla .\vec u,
\end{equation}
onde usamos a relação de Hubble $\vec u=\frac{\dot{a}}{a} \vec r$.
\par
Uma quest\~ao que surge imediatamente \'e se as modifica\c{c}\~oes introduzidas nas equa\c{c}\~oes (\ref{continuidade2}-\ref{poisson2}) n\~ao seriam pass\'{\i}veis de serem medidas em laborat\'orio.
Afinal, essas novas equa\c{c}\~oes representam uma modifica\c{c}\~ao da hidrodin\^amica usual. Mas, \'e preciso observar que a mudan\c{c}a consiste principalmente em adicionar a press\~ao divididada pela
velocidade da luz ao quadrado. Podemos realizar uma estimativa simples do efeito desta mudan\c{c}a. Essencialmente, ela consiste em introduzir uma "densidade de mat\'eria" dada por
$p/c^2$. Se considerarmos a press\~ao no fundo de uma fossa submarina, digamos a 10.000 metros, essa press\~ao seria dada por $10^9 N/m^2$, isto implicaria em acrescentar uma densidade de mat\'eria
da ordem de $10^{-9} kg/m^3$, o que corresponde a modificar a densidade da \'agua por um fator de $10^{-12}$. Isto equivale \`a ordem de grandeza dos efeitos relativistas
para os sistemas usuais que encontramos no laborat\'orio. No entanto, efeitos t\'{\i}picos de uma hidrodin\^amica relativista (que \'e mimetizado, de uma certa forma, pela teoria neo-Newtoniana)
podem aparecer em colis\~oes de \'{\i}ons pesados \cite{ollitrault}.

\subsection{Universo estático e o papel de $\Lambda$ na cosmologia neo-Newtoniana}

As mais diversas observações indicam que a dinâmica do Universo deve ser muito próxima- talvez exatamente- ao que é fornecida pelo modelo $\Lambda CDM$, brevemente descrito na primeira seção. Na verdade, o modelo $\Lambda CDM$ é o resultado de uma modificação na equação de Einstein que inclui a constante $\Lambda$. Não vamos discutir aqui a estrutura tomada pela equação de Einstein, mas sim suas soluções. Nesse caso, após assumirmos que a gravitação relativista está acoplada ao termo $\Lambda$, as equações (\ref{friedmann1}-\ref{friedmann3}) tornam-se
\begin{eqnarray}
\frac{\dot{a}^{2}}{a^{2}}+\frac{k\,c^2}{a^{2}}=\frac{8 \pi G \rho}{3}+\frac{\Lambda c^{2}}{3} \\ \frac{2\ddot{a}}{a}+\frac{\dot{a}^{2}}{a^{2}}+\frac{k\,c^{2}}{a^{2}}=-\frac{8 \pi G p}{c^{2}}+\Lambda c^{2} \\
\label{lambdaace}\frac{\ddot{a}}{a}=- \frac{4 \pi G}{3}\left( \rho + \frac{3 p}{c^{2}} \right)+\frac{\Lambda c^{2}}{3}. 
\end{eqnarray}
Efetivamente, podemos notar que as equações acima podem ser obtidas a partir de (\ref{friedmann1}-\ref{friedmann3}) se assumirmos que a densidade e a pressão podem ser reinterpretadas através da substituições: 
\begin{eqnarray}
\rho\rightarrow \rho+\frac{\Lambda c^{2}}{8\pi G}\hspace{2cm}p\rightarrow p-\frac{\Lambda c^{4}}{8\pi G}.
\label{subs}
\end{eqnarray}
\par
Vamos assumir agora a validade destas redefini\c c\~oes e modificar as equações da cosmologia neo-Newtoniana segundo (\ref{subs}). Por exemplo, a equação neo-Newtoniana (\ref{neoNewtace}) para a aceleração torna-se algebricamente idêntica a sua versão relativística (\ref{lambdaace}). As equações da continuidade e de Euler são invariantes sob a transformação (\ref{subs}), uma vez que envolvem combinações $\rho+\frac{p}{c^{2}}$. A principal diferença ao impormos as substituições (\ref{subs}) surge ao analisarmos a equação de Poisson neo-Newtoniana (\ref{poisson2}) resultando em:
\begin{eqnarray}
\nabla^{2} \Psi = 4 \pi G \left( \rho + \frac{3p}{c^{2}}\right) -\Lambda c^{2}.
\end{eqnarray}
\par
A equação acima admite como solução um potencial constante. De certa forma, esta equação está de acordo com a proposta feita por Einstein. Na defesa por um Universo estático, Einstein argumentou que um Universo com densidade constante deveria ter um potencial constante, de forma que a aceleração $\vec{a}=-\nabla \phi$ não exista. Essa proposta não \'e compat\'{\i}vel com a equação de Poisson (\ref{poisson}), o que o fez propor
\begin{equation}
\nabla^{2} \Psi +\lambda\Psi= 4 \pi G \rho,
\label{poissonEinstein}
\end{equation}
onde $\lambda$ seria a face newtoniana da constante cosmológica $\Lambda$. Um maneira moderna de reescrever (\ref{poissonEinstein}) é definir $\rho_v=\frac{\lambda \Psi}{4\pi G}$ tal que: $\nabla^{2} \Psi= 4 \pi G (\rho-\rho_v)$. Nesta proposta, a constante cosmológica torna-se solução do problema, no sentido
que contém o potencial gravitacional a ser determinado. Assim, o novo termo $\lambda$, na verdade, pode ser interpretado como uma densidade de matéria com propriedades anti-gravitacionais. Isso implica que o vácuo, ou seja, a ausência de matéria ordinária $(\rho=0)$, atuaria como fonte repulsiva de gravidade, que é o efeito associado à energia escura.

Mesmo que uma interpreta\c c\~ao da constante cosmologia n\~ao seja trivial dentro da cosmologia neo-Newtoniana (segundo as transforma\c c\~oes (\ref{subs})), devemos lembrar que n\~ao sabemos se o termo $\Lambda$ realmente existe na natureza. Na verdade, existe um forte argumento contra sua existência, o chamado {\it problema da constante cosmológica}. Se assumimos que $\Lambda$ é associada ao vácuo quântico, e que tal estado é descrito pela teoria quântica de campos até a escala de Planck, ent\~ao o valor teórico para a constante cosmológica é $\Lambda^{th}\sim M^4_{pl}$, onde $M_{pl}$ é a massa de Planck. Por outro lado, como as observa\c c\~oes nos dizem que o Universo é dominado pela energia escura, podemos inferir que seu valor observado é $\Lambda^{obs}\sim 10^{-120} M^4_{pl}$. Uma diferen\c ca de 120 ordens de magnitude! Uma alternativa a este cenário é admitir que a energia escura é descrita por um campo escalar (quintessência), descrita pela equa\c c\~ao de estado $p_{de}=w_{de} \rho_{de}$. Como resultado, as observa\c c\~oes indicam que $w_{de}\sim -1$. Dentro desta proposta, o estudo da energia escura e suas propriedades n\~ao apresentam os mesmo problemas que $\Lambda$, pois tratamos de um fluido com press\~ao, assim como necessitamos na cosmologia neo-Newtoniana. 

\section{Observações Finais}

Nosso objetivo nesse trabalho foi mostrar como tornou-se possível compreender a cosmologia sob um ponto de vista Newtoniano. Geralmente na ciência parte-se do "mais simples"\, para o "mais complexo". Mas não foi exatamente isso o que aconteceu com a cosmologia. A cosmologia Newtoniana surgiu apenas após os resultados relativísticos, que compreendem uma matemática bem mais complicada. Não há dúvidas que a RG fornece a melhor descrição para os fenômenos físicos que observamos no universo. Isso, inclusive, foi abordado pelo próprio McCrea ao deixar claro que eles não estavam sugerindo que a cosmologia Newtoniana eliminasse a necessidade do tratamento relativístico em grandes escalas \cite{McCrea1955}. No entanto, o que se observa é que a cosmologia Newtoniana pode simplificar o tratamento e fornecer as primeiras interpretações físicas do problema estudado. 
\par
Abordamos nesse trabalho desde os resultados de McCrea e Milne na década de 1930 até a formulação final de Harrison já na década de 1960, onde a pressão foi incorporada à cosmologia Newtoniana (sem nenhuma interpretação da RG), dando origem a cosmologia neo-Newtoniana. Em todo esse processo, percebemos que a crítica a essa formulação, que é uma abordagem simples para a cosmologia, sempre esteve muito presente e ainda é fonte de debate. Talvez, a questão principal seja, sobretudo, como estender resultados da teoria da gravitação Newtoniana local, que são muito bem testados em simples experimentos, para um novo laboratório como o cosmos onde as escalas de distâncias extrapolam o limite observado pelos melhores telescópios. Ao mesmo tempo, todo esse novo sistema, que deve compreender um espaço euclideano infinito, deve ser preenchido por matéria de maneira homogênea e isotrópica. Isso tudo gerou uma série de questionamentos que alcançaram, inclusive, o campo filosófico. 
\par
Todo o criticismo que circunda a cosmologia Newtoniana leva, em paralelo, a muitas alternativas para o problema cosmológico, que incluem alterações na gravitação Newtoniana para que esta se torne uma descrição eficaz da cosmologia. Para o problema de grandes escalas, talvez a primeira proposta tenha sido do astrônomo alemão Hugo Seeliger ainda no final do século $XIX$. Seeliger acreditava que ou a lei de força variando com o inverso do quadrado da distância estaria errada, ou ``a matéria total do universo deveria ser finita, ou ainda melhor, infinitas grandes partes do universo não deveriam ser preenchidas com massa ou uma densidade finita'' \footnote{``muss die Gesammtmaterie des Weltalls endlich sein oder genauer ausgedrückt, es dürfen nicht endlich grosse Theile des Raumes mit Masse von endlicher oder Dichtigkeit erfüllt sein'' \cite{Seeliger}.}. Isso fez Seeliger propor uma nova lei de força $F_S$ a partir de uma modificação da lei Newtoniana $F_N$ do tipo $F_S=F_N e^{-\lambda r}$, onde o fator de atenuação seria mais importante a partir de uma distância $r$ dada uma constante $\lambda$ \cite{Seeliger}. Como bônus, Seeliger descobriu que conseguiria resolver o problema do periélio de Mercúrio se adotasse $\lambda=3.8\times 10^{-7}m^{-1}$. Entretanto, esse valor, inevitavelmente, levaria a efeitos não observados nas órbitas dos outros planetas do sistema solar. Um série de outras propostas surgiram após Seerling e uma interessante discussão sobre esses avanços pode ser encontrada em \cite{norton1999}. 
\par
Aparentemente os resultados obtidos com a cosmologia neo-Newtoniana são compatíveis com a RG. Existem, porém, alguns trabalhos onde é verificado que a gravitação efetiva diverge da Newtoniana em escalas cosmológicas \cite{cluster}. Por outro lado, uma recente análise relativística para um particular modelo de matéria/energia chamado gás de Chaplygin generalizado\footnote{O gás de Chaplygin generalizado é um fluido, que se apresenta como candidato para o problema da matéria/energia escura, caracterizado por uma equação de estado $p=-\frac{A}{\rho^{\alpha}}$, onde $A$ e $\alpha$ são constantes.} \cite{winfried} confirmou resultados anteriores obtidos com a análise neo-Newtoniana \cite{neoN}. Isso nos dá mais uma indicação que mesmo para cenários mais complexos, e ainda sem solução, como o problema da matéria e energia escura, existe a possibilidade de se obter resultados confiáveis sem os altos custos matemáticos da relatividade geral.

A compara\c c\~ao entre cosmologias Newtonianas e relativísticas n\~ao se limita ao nível da expans\~ao de fundo do Universo, assim como focamos neste texto. Um próximo passo nessa discuss\~ao envolve o processo de forma\c c\~ao das estruturas cósmicas onde é necessário desenvolver um formalismo perturbativo para as grandezas físicas. Ainda temos muito o que aprender a partir das observa\c c\~oes da distribui\c c\~ao de matéria (galáxias reais), como também das simula\c c\~oes numéricas que tentam reproduzir tais dados. Curiosamente tais simula\c c\~oes, que utilizam a física Newtoniana, já atingem até mesmo escalas maiores do que o raio de Hubble. Mas afinal, seria a cosmologia Newtoniana ainda válida neste caso? Este é um tema que se encontra na vanguarda da pesquisa em cosmologia (veja recentes publica\c c\~oes em \cite{perturNewRel}). Pretendemos abordar com mais detalhes este ponto em uma futura comunica\c c\~ao.

\vspace{0.3cm}

{\bf Agradecimentos:} 
HESV agradece apoio do DFG através do projeto RTG 1620 ``Models of Gravity''. Agradecemos também ao CNPq e FAPES pelo apoio financeiro. A vers\~ao final desse manuscrito contou com as valiosas sugest\~oes de Antônio Brasil Batista e Oliver F. Piattella.


\begin{thebibliography}{99}
\addcontentsline{toc}{chapter}{\Large{Refer\^{e}ncias
Bibliogr\'{a}ficas}}

\bibitem{einstein1905}
A. Einstein, Sitz. Preuss. Akad. Wiss. Phys. {\bf 142} (1917), Ann. Phys. {\bf 69}, 436 (1922).

\bibitem{einstein1916}
A. Einstein, \emph{Annalen der Physik}, {\bf 49}, 769 (1916). 

\bibitem{kuhn}
T.S. Kuhn, {\bf A estrutura das revolu\c{c}\~oes cient\'{\i}ficas}, tradu\c{c}\~ao de Beatriz Vianna Boeira e Nelson Boeira, Editora Perspectiva, S\~ao Paulo (1982).

\bibitem{jean}
J. Eisenstaedt, {\bf Einstein et la relativit\'e g\'en\'erale}, \'Editions du CNRS, Paris (2002).

\bibitem{milne}
E. A. Milne, \emph{Quart. J. Math.} {\bf5}, 64 (1934).

\bibitem{milnemccrea}
E. A. Milne, W. H. McCrea, \emph{Quart. J. Math.} {\bf5}, 73 (1934). 

\bibitem{friedmann}
A. Friedmann, Z. Phys. {\bf 10}, 377 (1922).

\bibitem{hubble}
E. P. Hubble, Publ. Natu. Acad. Sci. {\bf 15}, 168 (1929).

\bibitem{lemaitre} G. Lema\^{\i}tre, Ann. Soc. Sci. de Bruxelles, {\bf 47}, 49 (1927).

\bibitem{bergh} S. van den Bergh, {\it Discovery of the Expansion of the Universe}, arXiv:1108.0709.

\bibitem{ademir}
J. A. S. Lima, V. Zanchin e  R. Brandenberger, MNRAS, {\bf 291}, L1-L4 (1997).

\bibitem{waga}
I. Waga, Revista Brasileira de Ensino de F\'{\i}sica,v. 27, n. 1, p. 157 - 173, (2005)

\bibitem{marcelo} M. B. Ribeiro, Boletim da Soc. Astronômica Brasileira, vol. 14, n $^{\underline{o}}$ 2, 34 (1994). 

\bibitem{Relatividade}
A. Einstein, {\bf Teoria da Relatividade Especial e Geral}, Contraponto, Rio de Janeiro (1999).

\bibitem{layzer}
D. Layzer, \emph{Astron. J}. {\bf59}, 258 (1954).

\bibitem{bondi}
H. Bondi, {\bf Cosmology}, Cambridge university press, Cambridge (1952).

\bibitem{mavrides} S. Mavrid\`es, {\bf L'Univers relativiste}, Masson et cie, Paris (1973).

\bibitem{norton1993}
J. Norton, PSA, vol. 2. East Alnsing: Philosophy of Science Association, pp. 412 (1992).

\bibitem{norton1995}
J. Norton, Philosophy of Science, {\bf 62}, 511 (1995).

\bibitem{malament}
D. Malament, Philosophy of Science, {\bf 62}, 489 (1995).
 
\bibitem{cartan}
E. Cartan, Annales Scientifiques de l'Ecole Normale Sup\'erieure {\bf 40}, 325-412, 1-25  (1923,1924). 

\bibitem{friedrichs}
K. Friedrichs, Mathematische Annalen {\bf98}, 566 (1927).

\bibitem{tipler}
F. J. Tipler, \emph{Am. J. Phys.} {\bf64}, 10 (1996).

\bibitem{landau}
L. Landau e E. Lifchitz, {\bf M\'ecanique des fluides}, \'Editions Mir, Moscou (1967).

\bibitem{mccrea1951}
W. H. McCrea, \emph{Proc. R. Soc. London} {\bf206}, 562 (1951). 

\bibitem{harrison}
E. R. Harrison, Ann. Phys (N.Y.) {\bf35}, 437 (1965).

\bibitem{whittaker}
E. T. Whittaker, \emph{Proc. Roy. Soc. A}, {\bf149}, 384 (1935). 

\bibitem{Ellis} S.W. Hawking e G.F.R. Ellis, {\bf Large scale structure of space-time}, Cambridge University press, Cambridge (1973).

\bibitem{ollitrault} J-Y. Ollitrault, Eur. J. Phys. {\bf 29}, 275 (2008).
  
\bibitem{McCrea1955} W. H. McCrea, AJ, {\bf 60}, 27 (1955).

\bibitem{Seeliger} H. von Seeliger, Astronomische Nachrichten, {\bf 137}, 129 (1895).

\bibitem{norton1999}
J. Norton, "The Cosmological Woes of Newtonian Gravitation Theory," in H. Goenner, J. Renn, J. Ritter and T. Sauer, eds., The Expanding Worlds of General Relativity: Einstein Studies, vol.7, Boston: Birkhauser, pp. 271-322 (1999).

\bibitem{cluster}
A. Shirata, T. Shiromizu, N. Yoshida e Y. Suto, Phys.Rev. D {\bf 71}, 064030 (2005). 

\bibitem{winfried}
J. C. Fabris, H. E. S. Velten e W. Zimdahl, Phys. Rev. D {\bf 81}, 087303 (2010). 

\bibitem{neoN}
J.C. Fabris, S.V.B. Gon\c calves, H.E.S. Velten e W. Zimdahl, Phys. Rev. D {\bf 78}, 103523 (2008).

\bibitem{perturNewRel} S. R. Green e R. M. Wald, Phys.Rev.D {\bf 85}, 063512 (2012); N. E. Chisari e M. Zaldarriaga, Phys.Rev.D {\bf 83}, 123505 (2011); S. Rasanen, Phys.Rev.D {\bf 81}, 103512 (2010).  

\end{thebibliography}
\end{document}